\documentclass[11pt]{article}
\usepackage[english]{babel}
\usepackage[latin1]{inputenc}
\usepackage{amsmath,amsfonts}

\title{Approachability of Convex Sets \\in Games with Partial Monitoring}
\author{Vianney Perchet\thanks{\'Equipe Combinatoire et Optimisation, FRE 3232 CNRS,
Universit\'e Pierre et Marie Curie - Paris 6, 4 place Jussieu, 75005
Paris. vianney.perchet@normalesup.org}}

\begin{document}
\maketitle
\newcounter{compteur}
\newtheorem{proposition}{Proposition}[section]
\newtheorem{theorem}[proposition]{Theorem}
\newtheorem{lemma}[proposition]{Lemma}
\newtheorem{corollary}[proposition]{Corollary}
\newtheorem{definition}[proposition]{Definition}
\newtheorem{remark}[proposition]{Remark}
\newtheorem{example}[proposition]{Example}

\begin{abstract}
We provide  a necessary and sufficient condition under which
 a convex set is approachable in a game with partial monitoring, i.e.\ where players
do not observe their opponents' moves but receive random signals. This condition is an extension of Blackwell's Criterion in the full monitoring framework, where players observe at least their payoffs. When our condition is fulfilled, we construct explicitly an approachability strategy, derived from  a strategy satisfying some \textsl{internal consistency} property in an
auxiliary game.

We also provide an example of a convex set, that is neither
(weakly)-approachable nor (weakly)-excludable, a situation that cannot occur in the full monitoring case.

We finally apply our result to describe an $\varepsilon$-optimal strategy  of  the
uninformed player in a zero-sum repeated game with incomplete
information on one side.

\bigskip

\textsl{Key Words :}   Repeated Games, Blackwell Approachability,
Partial Monitoring, Convex Sets, Incomplete Information
\end{abstract}
\normalsize

\section*{Introduction}

Blackwell\,\cite{BlackwellAnalogue} introduced the notion of
approachability in two-person (infinitely) repeated games with
vector payoffs in some Euclidian space $\mathbb{R}^d$, as an
analogue of Von Neumann's minmax theorem. A player can approach a
given set $E \subset \mathbb{R}^d$, if he can insure that, after
some stage and with a great probability, the average payoff will
always remain close to $E$. Blackwell\,\cite{BlackwellAnalogue}
proved that if both players observe their payoffs and $E$ satisfies
some geometric condition ($E$ is then called a $B$-set), then
Player\,1 can approach it. He also deduced that given a convex set
$C$ either Player\,1 can approach it or Player\,2 can exclude it,
\textsl{i.e.\ } the latter can approach the complement of a neighborhood  of
$C$. As Soulaimani, Quincampoix \& Sorin
\cite{AsSoulaimaniQuincampoixSorin} have recently proved that the
notions of $B$-set (in a given repeated game) and discriminating
domains (for a suitably chosen differential game) coincide.

\medskip

We consider the partial monitoring framework, where players do not
observe their opponent's moves but receive random signals. We
provide in section \ref{sectionPM} a necessary and sufficient
condition under which a convex set is  approachable. We also
construct  an approachability strategy derived from the construction
(following Perchet \cite{PerchetCalibrationALT}) of a strategy  that
has no internal regret (internal consistency in this framework has
been defined by  Lehrer \& Solan \cite{LehrerSolanPSE}, Definition
9).

Three classical results that hold in the full monitoring case do not extend to
 the partial monitoring  framework. Indeed, in a specific game introduced in section \ref{sectioncontreexample}, there exists a convex set  $C$  that is neither
approachable by Player 1   nor excludable by Player 2 (see  Theorem
3 in Blackwell\,\cite{BlackwellAnalogue}). Moreover, $C$ is not
approachable by Player 1 while every half-space that contains it is
approachable by Plater 1 (see Corollary 2 in
Blackwell\,\cite{BlackwellAnalogue}). Finally, $C$ is neither
weakly-approachable nor weakly-excludable (see
Vieille\,\cite{Vieille}). We recall that weak-approachability is a
weaker notion than approachability, also introduced by
Blackwell\,\cite{BlackwellAnalogue},  in finitely repeated games
(see Definition \ref{defweakapproach} in section
\ref{sectionapproach}).

\medskip

 Kohlberg \cite{Kohlberg} used the notion of approachability in order to construct optimal strategies of the uninformed
player, in the class of zero-sum repeated games with incomplete
information on one side (introduced by Aumann \&
Maschler\,\cite{AumannMaschler}). Our result can be used in this
framework to  provide a simple proof of the existence of a value in
the infinitely repeated game through the construction of an
$\epsilon$-optimal strategy of Player\,2.

\section{Approachability}\label{sectionapproach}

Consider a two-person game $\Gamma$ repeated in discrete time. At
stage $n \in \mathbb{N}$, Player\,1 (resp.\ Player\,2) chooses an
action $i_n \in I$ (resp.\ $j_n \in J$), where both sets $I$ and $J$
are finite. This generates a vector payoff   $\rho_n =
\rho(i_n,j_n) \in \mathbb{R}^d$ where $\rho$ is a mapping from $I \times J$ to
$\mathbb{R}^d$. Player\,1 does not observe $j_n$ nor $\rho_n$ but
receives   a random signal $s_n \in S$ whose law is $s(i_n,j_n)$
where $s$ is a mapping from  $I \times J$ to $\Delta(S)$  (the set
of probabilities over the finite set $S$). Player\,2 observes $i_n$,
$j_n$ and $s_n$. The choices of $i_n$ and $j_n$ depend only on the
past observations of the players and may be random.

\medskip

Explicitly, a strategy $\sigma$ of Player\,1  is a mapping from
$H^1$ to $\Delta(I)$ where $H^1 = \bigcup_{n \in \mathbb{N}} \left(I
\times S\right)^n$ is  the  set of finite histories
 available to Player\,1. After the
finite history $h_n^1 \in \left(I\times S\right)^{n}$,
$\sigma(h_n^1) \in \Delta(I)$ is the law of $i_{n+1}$. Similarly,  a
strategy $\tau$ of Player\,2 is a mapping from  $H^2=\bigcup_{n \in
\mathbb{N}}\left(I \times S \times J\right)^n$ to $\Delta(J)$.   A
couple of strategies $(\sigma,\tau)$ generates a probability,
denoted by $\mathbb{P}_{\sigma,\tau}$, over $\mathcal{H}=\left(I
\times S\times J \right)^{\mathbb{N}}$, the set of plays  embedded
with the cylinder $\sigma$-field.

The two functions $\rho$ and $s$ are extended multilinearly to
$\Delta(I) \times \Delta(J)$ by
$\rho(x,y)=\mathbb{E}_{x,y}\left[\rho(i,j)\right]\in \mathbb{R}^d$
and $s(x,y)=\mathbb{E}_{x,y}\left[s(i,j)\right]\in\Delta(S)$.

\medskip
The following notations will be used: for any sequence $a=\{a_m \in
\mathbb{R}^d\}_{m \in \mathbb{N}}$, the average of $a$ up to stage $n$ is denoted by  $\overline{a}_n:=\sum_{m =1}^n
a_m\big/n$ and for any set $E \subset \mathbb{R}^d$, the distance to $E$ is denoted by
$d_E(z):=\inf_{e \in E}\| z-e\|$, where $\|\cdot\|$ is the Euclidian norm.

\begin{definition}[Blackwell \cite{BlackwellAnalogue}]\label{defapproach}
\begin{itemize}
 \item[i)]{A closed  set $E \subset \mathbb{R}^d$ is approachable by Player\,1 if for every  $\varepsilon >0$, there exist
a strategy  $\sigma$ of Player\,1 and $N \in \mathbb{N}$ such that
for every strategy  $\tau$ of Player\,2 and every $n \geq N$:
\[ \mathbb{E}_{\sigma,\tau}\left[d_{E}(\overline{\rho}_n)\right]\leq \varepsilon \quad \mathrm{and} \quad \mathbb{P}_{\sigma,\tau}\left(\sup_{n \geq N}d_{E}(\overline{\rho}_n)\geq \varepsilon \right)\leq
 \varepsilon.
\]

Such a strategy $\sigma_{\varepsilon}$ is  called an
$\varepsilon$-approachability strategy of $E$.}
\item[ii)]{A set $E$ is excludable by Player\,2, if there exists  $\delta>0$ such that the complement of  $E^{\delta}$ is approachable by Player\,2, where $E^{\delta}=\{z \in \mathbb{R}^d; d_E(z) \leq \delta\}$.}
\end{itemize}
\end{definition}

In words, a set $E \subset \mathbb{R}^d$ is approachable by
Player\,1, if he can  insure  that  the average payoff  converges
almost surely to $E$, uniformly with respect to the strategies of
Player\,2. Obviously, a set $E$ cannot be both approachable by
Player\,1 and excludable by Player\,2.

\begin{definition}\label{defweakapproach}\begin{itemize}
 \item[i)]{A closed  set $E$ is weakly-approachable by Player\,1 if for every  $\varepsilon >0$, there exists $N \in \mathbb{N}$ such that for every $n \geq N$, there is some strategy  $\sigma_n$ of Player\,1  such that for every strategy  $\tau$ of Player\,2:
\[ \mathbb{E}_{\sigma_n,\tau}\left[d_{E}(\overline{\rho}_n)\right]\leq \varepsilon.\]}
\item[ii)]{A set $E$ is weakly-excludable by Player\,2, if there exists  $\delta>0$ such that the complement of  $E^{\delta}$ is weakly-approachable by Player\,2.}\end{itemize}
\end{definition}

We emphasize the fact that in the definition of weak-approachability, the strategy of Player\,1 might depend on $n$, the length of the game, which was not the case in the definition of approachability.
\subsection{Full monitoring case}
A game satisfies full monitoring if Player\,1 observes the moves of
Player\,2, thus if $S=J$ and $s(i,j)=j$.
Blackwell\,\cite{BlackwellAnalogue} gave a sufficient geometric
condition under which a closed set $E$ is approachable by Player  1.
He also provided  a full characterization for convex sets. Stating
his condition requires the following notations: $\Pi_{E}(z)
=\left\{e \in E; d_E(z)=\| z -e \|\right\}$ is the set of closest
points to $z \in \mathbb{R}^d$ in $E$, and $P^1(x) =\{\rho(x,y); y
\in \Delta(J)\}$ (resp.\ $P^2(y)=\{\rho(x,y); x\in \Delta(I)\}$) is
the set of expected payoffs compatible with $x \in \Delta(I)$
(resp.\ $y\in\Delta(J)$).

\begin{definition}\label{bset}
A closed set  $E$  of $\mathbb{R}^d$ is a  $B$-set, if for every $z
\in \mathbb{R}^d$, there exist $p \in \Pi_E(z)$ and
$x\left(=x(z)\right) \in \Delta(I)$  such that the hyperplane
through $p$ and perpendicular to $z-p$  separates $z$ from $P^1(x)$,
or formally:
\begin{equation}\label{blackwellcondition} \forall z \in \mathbb{R}^d, \exists p \in \Pi_E(z), \exists x \in \Delta(I),  \langle \rho(x,y) - p, z - p\rangle \leq 0, \quad \forall y \in \Delta(J).
\end{equation}
\end{definition}

Condition (\ref{blackwellcondition}) and therefore Theorem \ref{stratapprochparfaite} do not require that Player\,1 observes  Player\,2's moves, but only his own payoffs (which was Blackwell's assumption).

\begin{theorem}[Blackwell\,\cite{BlackwellAnalogue}]\label{stratapprochparfaite}
A $B$-set $E$ is approachable by Player\,1.

Moreover, consider the strategy $\sigma$ of Player\,1 defined by
$\sigma(h_n) = x(\overline{\rho}_n)$. Then   for every strategy
$\tau$ of Player\,2 and every $\eta>0$:
\begin{equation}
\mathbb{E}_{\sigma,\tau}[d^2_E(\overline{\rho}_n)]\leq \frac{4B}{n}
\quad \mathrm{and} \quad \mathbb{P}_{\sigma,\tau}\left(\sup_{n \geq
N}d_{E}(\overline{\rho}_n)\geq \eta\right)\leq \frac{8B}{\eta^2N},
\end{equation}
with $B=\sup_{i,j}  \|\rho(i,j)\|^2$.
\end{theorem}

For a closed convex set $C$,  a full characterization is available:
\begin{corollary}[Blackwell\,\cite{BlackwellAnalogue}]\label{repoussable}
A closed convex set $C \subset \mathbb{R}^d$ is approachable by
Player\,1 if and only if:
\begin{equation}\label{conditionconvexe}
P^2(y) \cap C \neq \emptyset, \quad  \forall y \in \Delta(J).
\end{equation}
\end{corollary}
Using a minmax argument, Blackwell\,\cite{BlackwellAnalogue} proved that  condition (\ref{conditionconvexe}) implies condition (\ref{blackwellcondition}), therefore the $B$-set $C$ is approachable by Player\,1. This characterization implies the following properties on convex sets:
\begin{corollary}[Blackwell\,\cite{BlackwellAnalogue}]\label{approachconvex}
\begin{enumerate}
\item A closed convex set $C$ is either  approachable by Player  1 or excludable by Player\,2.
\item A closed convex set $C$ is approachable by Player\,1 if and only if every half-space that contains $C$ is approachable by Player\,1.
\end{enumerate}
\end{corollary}
If condition (\ref{conditionconvexe}) is not fulfilled for some $y_0 \in \Delta(J)$, then (by the law of large numbers) Player\,2 just has to play accordingly to $y_0$ at each stage to exclude $C$. If every half-space that contains $C$ is approachable, then  $C$ is a $B$-set. Conversely any set that contains an approachable set is approachable.

\medskip

Blackwell also conjectured the following result on weak-approachability, proved by Vieille:
\begin{theorem}[Vieille\,\cite{Vieille}]\label{theovieille} A closed set is either weakly-approachable by Player\,1 or weakly-excludable by Player\,2.
\end{theorem}
Vieille\,\cite{Vieille} constructed a differential game $\mathcal{D}$ (in continuous time and with finite length) such that the finite repetitions of $\Gamma$ can be seen as a discretization of $\mathcal{D}$. The existence of the value for $\mathcal{D}$ implies the result.

\subsection{Partial monitoring case}\label{sectionPM}
The main objective of this section is to  provide a simple necessary and sufficient condition
under which  a convex set $C$ is approachable in the partial monitoring case.

Before stating it, we introduce the following notations: the vector
of probabilities over $S$ defined by $\mathbf{s}(y)=(s(i,y))_{i \in
I} \in \Delta(S)^I$ is called the flag generated by $y\in
\Delta(J)$. This flag is not observed by Player\,1 since if he plays $i \in I$  he only observes a signal $s$ which is the realization of the $i$-th  component of $\mathbf{s}(y)$. However, it is
theoretically the maximal information available to him about $y \in
\Delta(J)$. Indeed, Player 1 will never be able to distinguish between any two mixed action $y$ and $y'$ that generate the same flag, \textsl{i.e.\ } such that $\mathbf{s}(y)=\mathbf{s}(y')$.

Given a flag $\mu$ in $\mathcal{S}$, the range of $\mathbf{s}$,
$\mathbf{s}^{-1}(\mu)=\{y \in \Delta(J); \mathbf{s}(y)=\mu\}$ is the
set of mixed actions of Player\,2 compatible with $\mu$.
$P(x,\mu)=\{\rho(x,y); y \in \mathbf{s}^{-1}(\mu) \}$ is the set of
expected payoffs compatible with  $x \in \Delta(I)$ and  $\mu \in
\mathcal{S}$.

Our main result is:
\begin{theorem}\label{theogene}
A closed convex set $C \subset \mathbb{R}^d$ is approachable by
Player\,1 if and only if:
\begin{equation}\label{conditionsignaux}
\forall \mu \in \mathcal{S}, \exists x \in \Delta(I), P(x,\mu) \subset C.
\end{equation}
\end{theorem}
$P(x,\cdot)$ can be extended to $\Delta(S)^I$ (without changing
condition (\ref{conditionsignaux})) by defining, for every  $\mu \not \in \mathcal{S}$, either
$P(x,\mu)=\emptyset$ or $P(x,\mu)=P(x,\Pi_{\mathcal{S}}(\mu))$,
where $\Pi_{\mathcal{S}}(\cdot)$ is the projection onto
$\mathcal{S}$.

\medskip

 In the full monitoring
case, condition\,(\ref{conditionsignaux}) is exactly condition
(\ref{conditionconvexe}). Indeed, if Player 1 observes Player 2's
action then $S=J$, $\mathcal{S}=\{(y,\dots,y) \in \Delta(J)^I; y \in
\Delta(J)\}$ and given $\mathbf{y}=(y,\dots,y)\in\mathcal{S}$,
$P(x,\mathbf{y})=\{\rho(x,y)\}$. Condition (\ref{conditionsignaux})
implies that for every $y \in \Delta(J)$ there exists $x \in
\Delta(I)$ such that $\rho(x,y) \in C$, or equivalently $P^2(y) \cap
C \neq \emptyset$.

\medskip
An other important result is that  Corollary \ref{approachconvex} and Theorem \ref{theovieille}
do not extend:
\begin{proposition}\label{theoexample}
\begin{enumerate}
\item There exists a closed convex set that is neither approachable by Player\,1 nor excludable by
Player\,2
\item An half-space is either approachable by Player\,1 or excludable by
Player\,2
\item There exists a closed convex set that is not approachable by Player\,1 while every half-space that contains it is approachable by Player 1
\item There exists a closed convex set that is neither weakly-approachable by Player\,1 nor weakly-excludable by Player\,2.
\end{enumerate}
\end{proposition}

As said in the introduction, the proof of  Theorem \ref{theogene} relies on the construction of a
strategy that has no internal regret in an auxiliary game with partial monitoring.

\section{Internal regret with partial monitoring}
Consider the following two-person repeated game $\mathcal{G}$ with partial monitoring. At
stage $n \in \mathbb{N}$, we denote by $x_n \in \Delta(I)$ and $y_n \in \Delta(J)$ the mixed actions chosen by Player\,1 and Player\,2 (\textsl{i.e.\ } the laws of $i_n$ and $j_n$). As before, we denote by  $s_n$ the signal observed by Player\,1, whose law is the $i_n$-th coordinate of $\mu_n=\mathbf{s}(j_n)$.

Although payoffs are unobserved, given a flag $\mu \in \Delta(S)^I$
and $x \in \Delta(I)$, Player\,1 evaluates his payoff through
$G(x,\mu)$ where $G$ is a continuous map from $\Delta(I) \times
\Delta(S)^I$ to $\mathbb{R}$, not  necessarily linear.

\medskip

In the full monitoring framework, Foster \&
Vohra\,\cite{FosterVohraCalibratedLearningCorrelatedEquilibrium}
defined internally consistent strategies (or strategies that have no
internal regret) as follows: Player\,1 has asymptotically no
internal regret if for every $i \in I$, either the action $i$ is a
best response to his opponent's empirical distribution of actions on
the set of stages where he actually played $i$, or the density of
this set (also called the frequency of the action $i$) converges to
zero.

In our framework, $G$ is not linear so every action $i \in I$ (or
the Dirac mass on $i$) might never be a best response; best responses are indeed elements of $\Delta(I)$. Thus if we
want to define internal regret, we cannot distinguish the stages as a
function of the actions actually played (\textsl{i.e.\ } $i_n \in I$) but as a function of the laws of the actions (\textsl{i.e.\ } $x_n \in
\Delta(I)$).

We consider strategies described as follows: at stage $n$ Player\,1
chooses (at random) a law $x(l_n)$ in a finite set
$\{x(l)\in\Delta(I);l\in L\}$ and given that choice, $i_n$ is drawn
accordingly to $x(l_n)$; $l_n$ is called the type of the stage $n$.

\medskip

We denote by $N_n(l)=\{1 \leq m \leq n; l_m=l\}$ the set of stages
(before the $n$-th) of type $l$  and for any sequence $a=\{a_m \in
\mathbb{R}^d\}_{m \in \mathbb{N}}$, $\overline{a}_n(l)=\sum_{m \in
N_n(l)}a_m/|N_n(l)|$ is the average of $a$ on $N_n(l)$.

\begin{definition}\label{defiregret}
For every $n \in \mathbb{N}$ and every $l \in L$, the internal
regret of type $l \in L$ at stage $n$ is
\[\mathcal{R}_n(l)=\sup_{x\in\Delta(I)}\left[G(x,\overline{\mu}_n(l))-G(x(l),\overline{\mu}_n(l))\right],\]
where  $\overline{\mu}_n(l)$ is the unobserved average flag on $N_n(l)$.

A strategy $\sigma$ of Player\,1 is $(L,\varepsilon)$-internally
consistent  if for every strategy $\tau$ of Player\,2:
\[ \limsup_{n \to +\infty} \frac{|N_n(l)|}{n}\bigg(\mathcal{R}_n(l) - \varepsilon\bigg) \leq 0, \quad \forall l \in L, \quad \mathbb{P}_{\sigma,\tau}\mathrm{-as}.\]
\end{definition}
The set $L$ is assumed to be finite, otherwise there would exist
trivial strategies such that the frequency of every $x(l)$ converges
to zero. In words, if $\sigma$ is an $(L,\varepsilon)$-internally consistent strategy then  either $x(l)$ is an $\varepsilon$-best response to $\overline{\mu}_n(l)$, the
unobserved average flag on $N_n(l)$, or this set has a very small density.

\begin{theorem}[Lehrer \& Solan\cite{LehrerSolanPSE}; Perchet\,\cite{PerchetCalibrationALT}]\label{MICexistence}
For every $\varepsilon >0$, there
exist a finite set $L$ and a $(L,\varepsilon)$-internally
consistent strategy $\sigma$ such that for every strategy $\tau$ of Player 2:
\[
\mathbb{E}_{\sigma,\tau}\left[\sup_{l \in
L}\frac{|N_n(l)|}{n}\bigg(\mathcal{R}_n(l)-\varepsilon\bigg)\right]=
O\left(\frac{1}{\sqrt{n}}\right) \quad \mathrm{and}
\]
\[\forall \eta>0, \mathbb{P}_{\sigma,\tau}\left(\exists n \geq N, l \in L,\frac{|N_n(l)|}{n}\bigg(\mathcal{R}_n(l)-\varepsilon\bigg)>\eta \right)\leq O\left(\frac{1}{\eta^2N}\right).\]
\end{theorem}

\section{Proofs of the main results}
This section is devoted to the proofs of the theorems stated in the
previous section.
\subsection{Proof of Theorem \ref{theogene}}
 Let $C$ be a convex set such that for
every $\mu \in \Delta(S)^I$ there exists $x_{\mu} \in \Delta(I)$
such that $P(x_{\mu},\mu) \subset C$. Given $\varepsilon >0$, we are going to construct an $\varepsilon$-approachability strategy in $\Gamma$
based  on  an $(L,\epsilon)$-internally
consistent strategy in some auxiliary game $\mathcal{G}$, where the evaluation function $G$ is defined by:
\[
G(x,\mu)=  - \sup_{y \in
\mathbf{s}^{-1}(\mu)}d_C\left(\rho(x,y)\right) \] if
$\mu \in \mathcal{S}$. If $\mu \notin \mathcal{S}$, then  $G(x,\mu)=G\left(x,\Pi_{\mathcal{S}}(\mu)\right)$ where
$\Pi_{\mathcal{S}}$ is the projection onto $\mathcal{S}$.

\textbf{Sufficiency:} Any strategy in the auxiliary game
$\mathcal{G}$ naturally defines a strategy in the original game
$\Gamma$. The main idea of the proof is quite simple: given
$\varepsilon>0$, consider the finite family $\{x(l);l \in L\}$  and
the  $(L,\varepsilon)$-internally consistent strategy $\sigma$ of
Player\,1 given by Theorem \ref{MICexistence}. Then for every $l \in
L$, either $|N_n(l)|/n$ is very small, or $\mathcal{R}_n(l) \leq
\varepsilon$. In that last case, the definition of $G$ implies that
$\overline{\rho}_n(l)$  is $\varepsilon$-close to $C$. Since
\begin{equation}\label{convexpayoff}\overline{\rho}_n=\sum_{l \in
L}\frac{|N_n(l)|}{n}\overline{\rho}_n(l),\end{equation}
 $\overline{\rho}_n$ is  a convex combination of terms
that are $\varepsilon$-close to $C$. Since $C$ is convex,
$\overline{\rho}_n$ is also close to $C$.

\medskip

Formally, let $\sigma$ be a  $(L,\varepsilon)$-internally
consistent strategy of Player\,1 given by Theorem \ref{MICexistence}.
For every $\theta
>0$, there exists $N^1 \in \mathbb{N}$ such that for any strategy
$\tau$ of Player\,2:
\begin{equation}\label{preuve1}
 \mathbb{P}_{\sigma,\tau}\left(\forall n \geq N^1, \sup_{l \in L}\frac{|N_n(l)|}{n}\bigg(\mathcal{R}_n(l)-\varepsilon\bigg)\leq\theta\right)\geq 1- \theta.
\end{equation}
Recall that for any $\mu \in \Delta(S)^I$ there exists
$x_{\mu} \in \Delta(I)$ such that $P(x_{\mu},\mu)\subset C$, therefore
$\sup_{z \in \Delta(I)}G(z,\mu)=G(x_{\mu},\mu)=0$ and
\[\mathcal{R}_n(l)=\sup_{y \in \mathbf{s}^{-1}(\overline{\mu}_n(l))}d_C\left(\rho(x(l),y)\right)\geq d_C\bigg(\rho\big(x(l),\overline{\jmath}_n(l)\big)\bigg),\]
because $\mathbf{s}(\overline{\jmath}_n(l))=\mu_n(l)$ by linearity
of $\mathbf{s}$.

The random variables  $l_n$ and $j_n$ are independent (given the finite histories) and so are  $i_n$ and $j_n$ given $l_n$. Thus Hoeffding-Azuma~\cite{Azuma,Hoeffding}'s inequality for sums of
bounded martingale differences implies that
$\rho\left(x(l),\overline{\jmath}_n(l)\right)$ is asymptotically close to
$\overline{\rho}_n(l)$.  Explicitly, for every $\theta >0$, there
exists $N^2 \in \mathbb{N}$ (independent of $\sigma$ and $\tau$) such that:
\begin{equation}\label{preuve2}\mathbb{P}_{\sigma,\tau}\left(\forall n \geq N^2, \exists l \in L, \frac{|N_n(l)|}{n}\big| \overline{\rho}_n(l)-\rho\left(x(l),\overline{\jmath}_n(l)\right)\big|\leq \theta \right) \geq 1-\theta.
\end{equation}
Equations (\ref{preuve1}) and (\ref{preuve2}) imply that for every $n \geq
N=\max\{N^1,N^2\}$ and every $l \in L$, with
probability at least $1-2 \theta$:
\[\frac{|N_n(l)|}{n}\big( d_C\left(\overline{\rho}_n(l)\right)-\varepsilon\big) \leq 2\theta. \]

Since $C$ is a convex set, $d_C(\cdot)$ is convex, thus for any
strategy $\tau$ of Player\,2, with
$\mathbb{P}_{\sigma,\tau}$-probability at least $1-2\theta$, for
every  $n \geq N$:
\[d_C(\overline{\rho}_n)\leq \sum_{l \in L}\frac{|N_n(l)|}{n}d_C(\overline{\rho}_n(l))\leq 2 L\theta +\varepsilon, \]
and $C$ is approachable by Player\,1.

\medskip

\textbf{Necessity:} Conversely, assume that there exists $\mu_0 \in \Delta(S)^I$ such
that for all $x \in \Delta(I)$, there is some $y (=y(x)) \in
\mathbf{s}^{-1}(\mu_0)$ such that $d_C\left(\rho(x,y)\right)>0$.
Since $\Delta(I)$ is compact,  we can assume that there exists
$\delta>0$ such that $d_C\left(\rho(x,y(x))\right)\geq\delta$.

Let $\mathcal{T}_0$ be the subset  of strategies of Player\,2 that
generate at any stage the same flag $\mu_0$ (explicitly, a strategy
$\tau$ belongs to $\mathcal{T}_0$  if for every finite history
$h^2_n$, $\tau(h^2_n) \in \mathbf{s}^{-1}(\mu_0)$). Recall that a
strategy $\sigma$ of Player\,1 depends only on his past actions and
on the signals he received. Since at any stage, two strategies
$\tau$ and $\tau'$ in $\mathcal{T}_0$ induce the same laws of
signals,  the couples $(\sigma,\tau)$ and $(\sigma, \tau')$ generate
the same probability on the infinite sequences of moves of
Player\,1. Therefore
$\mathbb{E}_{\sigma,\tau}\left[\overline{\imath}_n\right]=\mathbb{E}_{\sigma,\tau'}\left[\overline{\imath}_n\right]:=\overline{x}_n$
is independent of $\tau$.

\medskip

For every $n \in \mathbb{N}$, define the strategy $\tau_n$ in
$\mathcal{T}_0$  by $\tau_n(h)=y(\overline{x}_n)$, for all finite
history $h$. Since $d_C(\cdot)$ is convex, by Jensen's inequality
\[\mathbb{E}_{\sigma,\tau_n}\left[d_C\left(\overline{\rho}_n\right)\right]\geq d_C\left(\mathbb{E}_{\sigma,\tau_n} \left[\overline{\rho}_n\right]\right).\]
Since $j_m$ is independent of the history $h_{m-1}$:
\[\mathbb{E}_{\sigma,\tau_n}\left[\rho(i_m,j_m)\big|h_{m-1}\right]=\mathbb{E}_{\sigma,\tau_n}\left[\rho(i_m,y(\overline{x}_n))\big|h_{m-1}\right]\] hence by linearity of $\rho(\cdot,y(\overline{x}_n))$,
\[\mathbb{E}_{\sigma,\tau_n}\left[\rho(i_m,j_m)\big|h_{m-1}\right]=\rho\left(\mathbb{E}_{\sigma,\tau_n}\left[i_m\big|h_{m-1}\right],y(\overline{x}_n)\right).\]
Therefore $\mathbb{E}_{\sigma,\tau_n}
\left[\overline{\rho}_n\right]=\rho(\overline{x}_n,y(\overline{x}_n))$.
Consequently
\[\mathbb{E}_{\sigma,\tau_n}\left[d_C\left(\overline{\rho}_n\right)\right]\geq d_C\left(\mathbb{E}_{\sigma,\tau_n} \left[\overline{\rho}_n\right]\right)= d_C\left(\rho(\overline{x}_n,y(\overline{x}_n))\right)\geq \delta\]
and for any strategy $\sigma$ of Player\,1 and any stage $n \in
\mathbb{N}$, Player\,2 has a strategy such that the expected average
payoff is at a distance greater than  $\delta >0$ from $C$. Thus  $C$
is not approachable by Player\,1.
\begin{remark}
The fact that $C$ is a convex set is crucial in both parts of the proof. In the  sufficient part, it would otherwise be possible that $\overline{\rho}_n(l) \in C$ for every $l \in L$, while $\overline{\rho}_n \notin C$. In the  necessary part, the counterpart could happen: $d_C\left(\mathbb{E}\left[\overline{\rho}_n\right]\right)\geq \delta$ while $\mathbb{E}\left[d_C(\overline{\rho}_n)\right]=0$.
\end{remark}

\begin{remark}\label{remarkrate}
The $\varepsilon$-approachability strategy constructed relies on a
$(L,\varepsilon)$-internally consistent strategy, so one can easily
show that:
 \[
\mathbb{E}_{\sigma,\tau}\left[d_C\left(\overline{\rho}_n\right)\right]=
\varepsilon+     O\left(\frac{1}{\sqrt{n}}\right) \quad
\mathrm{and}
\]
 \[
\mathbb{P}_{\sigma,\tau}\left(\exists n \geq N,
d_C\left(\overline{\rho}_n\right)-\varepsilon>\eta\right)\leq
O\left(\frac{1}{\eta^2N}\right).
\]
\end{remark}

\begin{corollary}\label{doublingtrick}
There exists $\sigma$ a strategy of Player\,1 such that for every $\eta >0$, there exists $N \in \mathbb{N}$ such that for every strategy $\tau$ of Player\,2 and $n \geq N$, $\mathbb{E}_{\sigma,\tau}\left[d_C\left(\overline{\rho}_n\right)\right] \leq \eta$.
\end{corollary}
The proof is rather classical and relies on a careful concatenation of $\varepsilon_k$-approachability strategies (where the sequence $(\varepsilon_k)_{k \in \mathbb{N}}$ decreases towards 0) called \textsl{doubling trick} (see e.g.\ Sorin~\cite{SorinSupergames}, Proposition 3.2). It is therefore omitted.

\subsection{Proof of Proposition \ref{theoexample}}\label{sectioncontreexample}
In the proof of Theorem \ref{theogene}, we have shown that if a
convex set is not approachable by Player\,1 then for any of his strategy
 and any $n \in \mathbb{N}$, Player\,2 has a
strategy $\tau_n$ such that $\overline{\rho}_n$ is at,  at
least, $\delta$ from $C$. It does not imply that $C$ is excludable by Player\,2; indeed this  would require that $\tau_n $ does not depend on $\sigma$ nor $n$. The proof of Proposition \ref{theoexample} relies mainly on the study of the following example.

\medskip
\textbf{Proof of Proposition \ref{theoexample}.}
 Consider the following matrix two-person repeated game where Player\,1 (the row player) receives no signal and his one-dimensional payoffs   are defined by :
\begin{tabular}{c|c|c|}
\multicolumn{1}{c}{}&\multicolumn{1}{c}{$L$}&\multicolumn{1}{c}{$R$}\\
\cline{2-3}
$T$&0&1\\
\cline{2-3}
$B$&-1&0\\
\cline{2-3}
\end{tabular}

\medskip
\textbf{$\mathbf{C:=[0;1/2]}$ is neither approachable nor excludable:}
The closed convex set $C:=[0;1/2]$ is obviously not approachable by
Player\,1 (otherwise Theorem \ref{theogene} implies that there exists
$x \in \Delta(I)$ such that $\rho(x,y)\in [0,1/2]$ for every $y \in
\Delta(J)$). More precisely, given  a strategy $\sigma$ of Player\,1, we define $\tau_n$ as follows: if $\overline{x}_n$ (the expected
frequency of $T$ up to stage $n \in \mathbb{N}$ --- it does not depend on Player\,2's strategy) is smaller than $1/4$, then $\tau_n$ is the strategy that always plays  $L$,
otherwise that always plays $R$. Then the law of large numbers implies that, for $n$ big enough,
$\mathbb{E}_{\sigma,\tau_n}\left[d_C\left(\overline{\rho}_n\right)\right]$ is arbitrarily close to $1/4$.

\medskip

It remains to show that Player\,2 cannot exclude $C$. We
prove this by constructing a strategy $\sigma$ of Player\,1 such that
the average payoff is  infinitely often close to 0:
 $\sigma$ is played in blocks and the length of the $p$-th block is $p^{2p+1}$. On odd blocks, Player\,1 plays $T$ while on even blocks he plays $B$. At the end of the block $p$, the average payoff is at most $1/p$ if it is an odd block and at least $-1/p$ otherwise. Hence on two consecutive blocks (the $p$-th and the $p+1$-th) there is at least one stage such that the average payoff is at a distance smaller than $1/p$ to $\{0\}$. Therefore $\{0\}$ and $C$ (since it contains $\{0\}$) is not excludable by Player\,2.

\medskip

\textbf{An half-space is either approachable by Player\,1 or excludable by Player\,2:}
 Let $E$ be an half-space not approachable by Player\,1. Then there exists $\mu_0 \in \Delta(S)^I$ such that, for every $x \in \Delta(I)$, $P(x,\mu_0) \not \subset E$. This implies that there exists $\delta >0$ such that  $\inf_{x \in \Delta(I)} \sup_{y \in \mathbf{s}^{-1}(\mu_0)} d_E\left(\rho(x,y)\right)\geq \delta>0$ and therefore for every $x \in \Delta(I)$, there exists $y \in \Delta(J)$ such that $\rho(x,y)$ is in the complement of $E^{\delta}$ which is convex, since $E$ is an half-space. Blackwell's result applies for Player\,2 (since we assumed he has full monitoring), so he can approach the complement of $E^{\delta}$ and exclude $E$.

 \medskip

\textbf{$\mathbf{C}$ is not approachable by Player\,1 while every half-space that contains it is:}
An half-space that contains $C$ contains either $(-\infty,0]$ or $[0,+\infty)$ which are approachable by, respectively, always playing $T$ or always playing $B$.

\medskip

\textbf{$\mathbf{C}$ is neither weakly-approachable by Player\,1 nor weakly excludable by Player\,2 :} we proved that for every strategy $\sigma$ of Player\,1 and every $n \in \mathbb{N}$ big enough, Player\,2 has a strategy $\tau_n$ such that $\mathbb{E}_{\sigma,\tau_n}\left[d_C(\overline{\rho}_n)\right]=1/2$. Hence $C$ is not weakly approachable.

\medskip

Conversely, let $\tau$ be a strategy of Player\,2 in the game repeated $2n$ times
(where $n$ is large enough) and $M \in \mathbb{N}$ be any integer. Consider the strategy $\sigma$ of Player\,1 that consists
in playing $T$ during the first $n$ stages. Since $\overline{\rho}_n$, the average payoff after
those $n$ stages, belongs to $[0; 1]$, there exists an integer $k_1 \in \{1,\dots,M\}$ such that
$\overline{\rho}_n$ belongs to $[\frac{k_1-1}{M}; \frac{k_1}{M}]$ with $\mathbb{P}_{\sigma,\tau}$-probability at least $\frac{1}{M}$. Note that,
given $\tau$ , Player\,1 can compute this $k$.

Assume that, from stage $n+1$ on,
the strategy $\sigma$ dictates to play i.i.d action $B$ with probability $\frac{k_1}{M}$ and action $T$ with
probability $1-\frac{k_1}{M}$. If $n$ is large enough, the probability that the average
payoff between stages $n+1$ and $2n$ belongs to $[-\frac{k_1}{M}-\frac{1}{M}; 1-\frac{k_1}{M}+\frac{1}{M}]$ is close to
one (say bigger than $1/2$, this is again a direct consequence of the law of large number). Therefore, this strategy $\sigma$ ensures that with $\mathbb{P}_{\sigma,\tau}$-probability at
least $\frac{1}{2M}$ the average payoff over the $2n$ stages belongs to $[-\frac{1}{M};\frac{1}{2}+\frac{1}{2M}]$.

\medskip

Denote by $\left(C^{2/M}\right)^c$ the complement of the $\frac{2}{M}$-neighborhood of $C$. Given a strategy $\tau$ of Player\,2 and an integer $n$ big enough, the strategy $\sigma$ we described ensures that
$\mathbb{E}_{\sigma,\tau}\left[d_{\left(C^{2/M}\right)^c}\left(\overline{\rho}_{2n}\right)\right] \geq \frac{1}{2M^2}$. Therefore, for every $M \in \mathbb{N}$, $\left(C^{2/M}\right)^c$ is not weakly-approachable by Player\,2 hence $C$ is not weakly-excludable.

\bigskip

The strategy $\sigma$ we described can be easily made independent of $\tau$ by, for example, choosing $k_1 \in \{1,\dots,M\}$ at random; indeed, this would imply that  $\mathbb{E}_{\sigma,\tau}\left[d_{\left(C^{2/M}\right)^c}\left(\overline{\rho}_{2n}\right)\right] \geq \frac{1}{2M^3}$. $\hfill \Box$

\bigskip

These results hold if one chooses $C_3:=[0;1/3]$ instead of $[0;1/2]$. In fact, it only remains to prove that $C_3$ is not weakly-excludable by Player\,2. Consider the game repeated $3n$ times and the strategy $\sigma$, defined by block of size $n$, that plays  on the first block always $T$, on the second block i.i.d. action $B$ with probability $\frac{k_1}{M}$. The average payoff on those two block belongs to a small neighborhood of $[0;1/2]$, hence to some $[\frac{k_2-1}{M},\frac{k_2}{M}]$ (where $k_2 \leq \frac{M}{2}$) with probability at least $\frac{1}{M}$. Assume that on the third block Player\,1 plays i.i.d action $B$ with probability $\frac{2k_2}{M}$ then the average payoff over the three blocks belongs to a small neighborhood of $[0;1/3]$ with probability at least $\frac{1}{(2M)^2}$. Therefore $C_3$ is not weakly excludable.

Since this proof can be generalized to any set $C_k=[0;\frac{1}{k}]$, even the  singleton $\{0\}$ is neither weakly-approachable nor weakly-excludable; we recall that  in the full monitoring framework all those convex sets are  approachable by Player\,1.

\subsection{Remarks on the counterexample}
Following Mertens, Sorin \& Zamir's notations \cite{MSZ} (see
Definition 1.2 p.\ 149),
  Player\,1 can
guarantee $\underline{v}$ in a zero-sum repeated game $\Gamma_{\infty}$ if
\[\forall \varepsilon >0, \exists \sigma_{\varepsilon}, \exists N \in \mathbb{N}, \mathbb{E}_{\sigma_{\varepsilon},\tau}\left[\overline{\rho}_n\right]\geq \underline{v}-\varepsilon, \forall \tau, \forall n \geq N,\]
where $\sigma_{\varepsilon}$ is a strategy of Player\,1, and $\tau$
any strategy of Player\,2. Player\,2 can defend $\underline{v}$ if:
\[\forall \varepsilon >0, \forall \sigma_{\varepsilon}, \exists \tau, \exists N \in \mathbb{N}, \mathbb{E}_{\sigma_{\varepsilon},\tau}\left[\overline{\rho}_n\right]\leq \underline{v}+\varepsilon, \forall n \geq N. \]
If Player\,1 can guarantee $\underline{v}$ and Player\,2 defend
$\underline{v}$, then $\underline{v}$ is the maxmin of
$\Gamma_{\infty}$. The minmax $\overline{v}$  is defined in a dual
way and $\Gamma_{\infty}$ has a value if
$\underline{v}=\overline{v}$.

\medskip

These definitions can be extended to the vector payoff framework: we
say that Player\,1 can guarantee a set $E$ if he can approach $E$:
\[\forall \varepsilon >0, \exists \sigma_{\varepsilon}, \exists N \in \mathbb{N}, \mathbb{E}_{\sigma_{\varepsilon},\tau}\left[d_E\left(\overline{\rho}_n\right)\right]\leq \varepsilon, \forall \tau, \forall n \geq N.\]

In the  counterexample  of the proof of Proposition \ref{theoexample},
Player\,1 cannot guarantee the convex set $C=\{0\}$ and Player\,2 cannot defend it since:
\[\exists \sigma, \forall \varepsilon >0, \forall \tau, \forall N \in \mathbb{N}, \exists n \geq N, \mathbb{E}_{\sigma,\tau}\left[d_C\left(\overline{\rho}_n\right)\right] \leq \varepsilon.\]
To keep the notations of zero-sum repeated game, one could say that
the game we constructed {\em has no maxmin}.

\medskip
Blackwell\,\cite{BlackwellAnalogue} also gave an example of a game (with
vector payoff) without maxmin in the full monitoring case. The main
differences between the two examples are:
 \begin{itemize}
 \item[i)]{ in the partial monitoring case this set can be convex (which cannot occur in the
full monitoring framework);}
\item[ii)]{the strategy of Player\,1 is such that the average payoff is infinitely often close to $C$. However, unlike  Blackwell's example, he does not know at which stages.}
    \end{itemize}

\section{Repeated game with incomplete information on one side, with partial monitoring}
Aumann \& Maschler\,\cite{AumannMaschler} introduced the class of
two-person zero-sum games with incomplete information on one side.
Those games are described as follows: Nature chooses  $k_0$ from a
finite set  of states $K$ according to some known probability $p \in
\Delta(K)$. Player\,1 (the maximizer) is informed about $k_0$  but
not Player\,2. At stage $m \in \mathbb{N}$, Player\,1 (resp.\
Player\,2) chooses $i_m \in I$ (resp.\ $j_m \in J$) and the payoff
is $\rho_m^{k_0}=\rho^{k_0}(i_m,j_m)$. Player\,1 observes $j_m$ and
Player\,2 does not observe $i_m$ nor $\rho_m$ but receives a signal
$s_m$ whose law is $s^{k_0}(i_m,j_m) \in \Delta(S)$. As in the
previous sections, we define
$\mathbf{s}^k(x)=\left(s^k(x,j)\right)_{j \in K}$, for every $x \in
\Delta(I)$.

\medskip
A strategy $\sigma$ (resp.\ $\tau$) of Player\,1 (resp.\ Player\,2)
is a mapping from $K\times \bigcup_{ \in \mathbb{N}} \left(I \times
J\times S\right)^m$ to $\Delta(I)$ (resp.\ from $\bigcup_{m \in
\mathbb{N}}\left(J \times S\right)^m$ to $\Delta(J)$). At stage
$m+1$, $\sigma(k,h_m^1)$ is the law of $i_{m+1}$ after the history
$h_m^1$ if the chosen state is $k$.

We define $\Gamma_1$  the one-shot game with expected payoff
$\sum_{k \in K}p^k \rho^k(x^k,y)$ and $\Gamma_{\infty}(p)$ the
infinitely repeated game. We  denote by $v_{\infty}(p)$ its  value,
if it exists (\textsl{i.e.\ } if both Player\,1 and Player\,2 can guarantee
it). Aumann \& Maschler \cite{AumannMaschler} (Theorem C, p.\ 191)
proved that $\Gamma_{\infty}(p)$ has a value and characterized it.
\medskip

Let us first  introduce the operator $\mathbf{Cav}$ and the
non-revealing game $D(p)$: for any function $f$ from $\Delta(I)
\times \Delta(J)$ to $\mathbb{R}$, $\mathbf{Cav}(f)(\cdot)$ is the
smallest (pointwise) concave function greater than $f$.

A profile of mixed actions $x=(x^k)_{k \in K} \in \Delta(I)^K$ is
non-revealing at $p \in \Delta(K)$ (and induces the flag $\mu \in \Delta(S)^J$) if the flag induced by $x$ is
independent of the state:
\[NR(p,\mu)=\left.\left\{x=(x^1,\dots,x^K) \in \Delta(I)^K  \right| \mathbf{s}^k(x^k)=\mu,  \forall k \mathrm{\ st \ }p^k >0\right\}.\]
We denote by  $NR(p)=\bigcup_{\mu \in \Delta(S)^J}NR(p,\mu)$ the set
of non-revealing strategies. For every $\mu \in \Delta(S)^J$,
$D(p,\mu)$ (resp.\ $D(p)$) is the one-stage game $\Gamma_1$ where
Player\,1 is restricted to $NR(p, \mu)$ (resp.\ $NR(p)$) and its
value is denoted by $u(p,\mu)$ (resp.\ $u(p)$), with
$u(p,\mu)=-\infty$ if $NR(p,\mu)=\emptyset$ (resp.\ $u(p)=-\infty$
if $NR(p)=\emptyset$).

\begin{theorem}[Aumann \& Maschler\,\cite{AumannMaschler}]
The game $\Gamma_{\infty}$ has a value defined  by
$v_{\infty}(p)=\mathbf{Cav}(u)(p)$.
\end{theorem}
\textbf{Proof.} Player\,1 can guarantee $u(p)$: indeed if $NR(p)
\neq \emptyset$, he just has to play i.i.d.\ an optimal strategy in
$NR(p)$ and otherwise $u(p)=-\infty$. Therefore, using the splitting
procedure (see Lemma 5.2 p.\ 25 in \cite{AumannMaschler}), Player 1 can
guarantee $\mathbf{Cav}(u)(p)$.

\medskip

It remains to show that Player\,2 can also guarantee $\mathbf{Cav}(u)(p)$. The function $\mathbf{Cav}(u)(\cdot)$ is concave and continuous,
therefore there exists $\mathbf{m}=(\mathbf{m}^1,\dots,\mathbf{m}^k)
\in \mathbb{R}^K$ such that $\mathbf{Cav}(u)(p)=\langle \mathbf{m},p
\rangle$ and $u(q)\leq \mathbf{Cav}(u)(q) \leq \langle \mathbf{m},q
\rangle$. Instead of constructing a strategy of Player\,2 that minimizes the expected payoff $\sum_{k \in K}p^k \overline{\rho}_n^k$, it is enough to construct a strategy such that each $\overline{\rho}_n^k$ is smaller than $\mathbf{m}^k$, for every state $k$ that has a positive probability accordingly to Player\,2's posterior.

Therefore, we consider an auxiliary two-person  repeated game with
vector payoff where at stage $n \in \mathbb{N}$, Player\,2 (resp.\
Player\,1) chooses $j_n$ accordingly to $y_n \in \Delta(J)$ (resp.\
$(i_n^1,\dots,i_n^K)$ accordingly to $(x_n^1,\dots,x_n^K)\in
\Delta(I)^K$). Player\,2 receives a signal $s_n$ whose law is
$s^{k_0}(i_n^{k_0},j_n)$ where $k_0$ is the true state. We denote by
$\mu_n=\mathbf{s}^{k_0}(x_n^{k_0})$ the expected flag of stage $n$.
The $k$-th component of the vector payoff $\rho_n$ is defined by
$\rho^k(i_n^k,j_n)$ if $\mu_n$ belongs to $\mathcal{S}^k$,  the
range of $\mathbf{s}^k$ and $-A:=-\max_{k \in K}\|
\rho^k\|_{\infty}$ otherwise\footnote{We use this notation, because
if $\mu_n$ is not in the range of $\mathbf{s}^k$, then Player\,2
knows that the true state is not $k$, and therefore does not need to
minimize the $k$-th component of the payoff vector}. Conversely, the
set of compatible payoffs given a flag $\mu \in \Delta(S)^J$, $y \in
\Delta(J)$ and a state $k$, is defined by:
\[ P^k(\mu,y)=\left\{\rho^k(x^k,y) \left| \mathbf{s}^k(x^k)=\mu\right.\right\} \mathrm{ \ if \ } \mu \in  \mathcal{S}^k, \mathrm{ \ otherwise\ }P^k(\mu,y) = \{-A\}, \]
and the set of compatible vector payoffs  is  $P(\mu,y)=\Pi_{k \in
K} P^k(\mu,y) \subset \mathbb{R}^K$.

If Player\,2 can approach $M = \{m\in\mathbb{R}^K; m^k\leq
\mathbf{m}^k, \forall k\in K\}=\mathbf{m}+\mathbb{R}_-^K$, then he
can guarantee $\mathbf{Cav}(u)(p)$. Theorem \ref{theogene} implies
that the convex set $M$ is approachable if and only if, for every
$\mu \in \Delta(S)^I$ there exists $y \in \Delta(J)$ such that
$P(\mu,y) \subset M$.

Hence it is enough to prove that this property holds. Assume the converse:
there exists $\mu_0 \in \Delta(S)^I$ such that for every $y \in
\Delta(J)$, $P(\mu_0,y)$ is not included in $M$.

We denote by $K(\mu_0)=\left\{k \in K; \mu_0 \in
\mathcal{S}^k\right\}$ the set of states that are compatible with
$\mu_0$: if Player\,2 observes $\mu_0$, then he knows that the true
state is in $K(\mu_0)$. For every $y \in \Delta(J)$ and $k \in
K(\mu_0)$, $\omega^k_0(y)=
\sup_{\mathbf{s}^k(x^k)=\mu_0}\rho^k(x^k,y)$ is the worst payoff for
Player\,2 in state $k$. The fact that $P(\mu_0,y)$ is not included
in $M$ implies that $\omega_0(y)=(\omega_0^k(y))_{k \in K(\mu_0)}$
does not belong to $M_0=\{m \in \mathbb{R}^{K(\mu_0)}; m^k \leq
\mathbf{m}^k, \forall k  \in K(\mu_0)\}$. Define the convex set:
\[W_0=\left\{\omega_0(y); y \in \Delta(J)\right\} + \mathbb{R}^{K(\mu_0)}_+ \bigcap B(0,A),\] with $B(0,A)$ the closed ball of radius $A$.  Obviously $W_0\cap M_0 = \emptyset$ and, by linearity of each $\rho^k$, $W_0$ is a compact convex set. So there exists a strongly separating hyperplane $H_0=\{\omega \in \mathbb{R}^{K(\mu_0)}; \langle \omega,q_0\rangle = b\}$ such that $\sup_{m \in M_0} \langle m,q_0 \rangle < \inf_{\omega \in W_0} \langle \omega, q_0\rangle$.
Every component of $q_0$ must be non-negative (since $M_0$ is
negatively comprehensive), therefore up to a normalization, we can
assume that $q_0$ belongs to $\Delta(K(\mu_0))$.

Define $W=W_0 \times \mathbb{R}^{K\backslash K(\mu_0)}$ and  $q \in
\Delta(K)$ by $q(k)=q_0(k)$ if $k \in K(\mu_0)$ and 0 otherwise.
Then,  $H=\{\omega \in \mathbb{R}^{K}; \langle \omega,q\rangle =
b\}$ strongly separates $M$ and $W$, therefore:
 \[\langle \mathbf{m}, q \rangle <
\min_{\omega \in W_0} \langle \omega, q\rangle =\min_{y \in
\Delta(J)} \max_{x \in NR(q,\mu_0)} \sum_{k \in K}q^k
\rho^k(x^k,y)=u(q,\mu_0)\leq u(q)\] and by  definition of
$\mathbf{m}$,   $u(q) \leq \langle \mathbf{m},q\rangle$ which is
impossible.

So $M$ is approachable by Player\,2, he can guarantee
$\mathbf{Cav}(u)(p)$ in $\Gamma_{\infty}(p)$ and
$v_{\infty}(p)=\mathbf{Cav}(u)(p)$. $\hfill \Box$

\bigskip

\textbf{Acknowledgments:} I deeply thank my advisor Sylvain Sorin
for its great help and numerous comments. I also acknowledge useful
comments from J\'er\^ome Renault and Gilles Stoltz and, of course,
thank them for pointing out the counter example to me.

\end{document}